\documentclass[11pt]{article}

\usepackage{epsfig}
\newcommand{\noi}{\noindent}

\def\be{\begin{equation}}
\def\ee{\end{equation}}
\def\bea{\begin{eqnarray}}
\def\eea{\end{eqnarray}}
\def\gm{\gamma_\mu}
\def\g5{\gamma_5}
\def\vep{\varepsilon}

\def\sl{\vec {\sigma_l}}
\def\s{\vec \sigma}
\def\ve{\vec \varepsilon}
\def\kh{\hat k}
\def\nh{\hat  \nu}

\def\al{\alpha}
\def\pa{\partial}
\def\Journal#1#2#3#4{{#1}\, #3,\, #2,\, #4\,}
\def\NPA{{\em Nucl. Phys.} A}
\def\NPB{{\em Nucl. Phys.} B}
\def\PRL{\em Phys. Rev. Lett.\,\,}
\def\PRC{{\em Phys. Rev.} C}
\def\PRD{{\em Phys. Rev.} D}
\def\PRW{{\em Phys. Rev.}\,\,}
\def\CJPB{{\em Czech. J. Phys.} B}
\def\RMP{\em Rev. Mod. Phys.\,\,}
\def\PLB{{\em Phys. Lett.} B}
\def\PR{{\em Phys. Rep.}\,\,}
\def\JPG{{\em J. Phys. G: Nucl. Part. Phys.}\,\,}
\def\NCA{{\em Nuovo Cimento}\,\,}
\def\FBS{{\em Few--Body Syst.}\,\,}
\def\PHA{{\em Physica} A}
\def\PTP{{\em Progress of Theoretical Physics}\,\,}
\def\EPJC{{\em Eur. Phys. J.} C}

\begin{document}

\title{\bf Anomalous Lagrangians and the radiative muon
capture in hydrogen}

\author{
J. Smejkal \\
{\small Institute of Nuclear Physics, Academy of
Sciences of the Czech Republic,} \\
{\small CZ-25068 \v{R}e\v{z}, Czechia,
      ~email: smejkal@ujf.cas.cz}
\vspace{0.5 cm} \\
E. Truhl\'{\i}k \\
{\small Institute of Nuclear Physics, Academy of
Sciences of the Czech Republic,} \\
{\small CZ-25068 \v{R}e\v{z}, Czechia, ~email: truhlik@ujf.cas.cz}
\vspace{0.5 cm} \\
F.C. Khanna \\
{\small Theoretical Physics Institute, Department of
Physics, University of Alberta,} \\
{\small Edmonton, Alberta, Canada, T6G 2J1} \\
 {\small and TRIUMF,} \\
 {\small 4004 Wesbrook Mall, Vancouver, BC, Canada, V6T 2A3,
      ~email: khanna@phys.ualberta.ca}
}
\maketitle


  \begin{abstract}
  \noi
  The structure of an anomalous Lagrangian of the $\pi\rho\omega
  a_1$ system is investigated within the hidden local
  $SU(2)_R\times SU(2)_L$ symmetry approach.  The interaction of
  the external electromagnetic and weak vector and axial--vector
  fields with the above hadron system is included.

   The Lagrangian of interest contains the anomalous Wess--Zumino
   term following from the well known
  Wess--Zumino--Witten action and six independent homogenous
terms. It is characterized by four constants that are to be
determined from a fit to the data on various elementary reactions. Present
data allows one to extract the constants with a good accuracy.

The homogenous part of the Lagrangian has been applied in the
  study of anomalous processes that could enhance the high
energy tail of the spectrum of photons, produced in the radiative
muon capture in hydrogen.
  It should be noted that  recently,
  an intensive search for such enhancement processes has been
  carried out  in the literature,  in an attempt
to resolve the so called \mbox{"$g_P$ puzzle":} an $\approx$ 50 \% difference
between the theoretical prediction of the value of the induced
  pseudoscalar constant $g_P$ and its value extracted from the high
  energy tail of the photon spectrum, measured in the
precision TRIUMF experiment.

  Here, more details on the studied material are presented
and new results, obtained by using the Wess--Zumino term,
are  provided.

\end{abstract}

\section{Introduction}
\label{intro}

The theory of the weak nuclear interaction aims to describe
nuclear phenomena induced by the external interaction, that is
mediated by the intermediate bosons $W^\pm$ and $Z^0$ of the
Standard Model \cite{DGH,W2}. At low and intermediate energies,
the  strangeness conserving semileptonic nuclear interaction
Hamiltonian is of the current-current form \cite{AD}-\cite{JDW0}
\be {\cal H}_W\,=\,-\frac{G_F}{\sqrt 2}\,cos\,
\theta_C\,\left[\,{\hat J}^a_{W,\,\mu}\,
l^*_\mu\,+\,h.c.\,\right]\,.  \label{HWN} \ee Here the weak
interaction constant $G_F/(\hbar c)^3\,\approx\,1.16637(1) \times
10^{-5}\,GeV^{-2}$ \cite{PDG}, the Cabibbo angle \cite{CA} $cos\,
\theta_C\,=\,0.9738(5)$ \cite{PDG}, $l_\mu$ is the lepton current
and the operator of the weak nucleon current is
\bea
{\hat J}^a_{W,\,\mu}(q_1)&=&{\hat J}^a_{V,\,\mu}(q_1)\,
+\,{\hat J}^a_{A,\,\mu}(q_1) \nonumber \\
&=&i\left(g_V(q^2_1)\gm\,-\,\frac{g_M(q^2_1)}
{2M}\,\sigma_{\mu\,\nu}\,q_{1\nu}\,-\,g_A(q^2_1)\gm\g5\,+\,
i\frac{g_P(q^2_1)}{m_l}q_{1\mu}\g5\right)\frac{\tau^a}{2}\,,
\label{ONWHC}
\eea
where $M(m_l)$ is the nucleon (lepton) mass and
$q_{1\mu}=p'_\mu-p_\mu$, where $p'_\mu(p_\mu)$ is the 4-momentum of the
final (initial) nucleon.

The least known of the form factors entering Eq.\,(\ref{ONWHC}) is
the induced pseudoscalar form factor $g_P(q^2_1)$. The presence of
this form factor in the weak nucleon current is a consequence of
the intimate relation between the strong and weak interaction
processes. The contribution of the pion pole to $g_P(q^2_1)$ is
\be g_P(q^2_1)\,=\,-2 g_{\pi NN} f_\pi m_l \Delta^\pi_F(q^2_1)\,,
\label{gP} \ee where $g_{\pi NN}$ is the pseudoscalar $\pi NN$
coupling constant
\footnote{Modern phenomenological nucleon-nucleon potentials use
$g_{\pi NN}\approx 13.0.$}, $f_\pi$ is the pion decay constant~%
\footnote{According to Ref.\,\cite{PDG}, ${\sqrt
2}f_\pi^+=130.7\pm0.1\pm0.36\,MeV$, thus providing $f_\pi^+=92.7\pm 0.3\,MeV$;
let us note, however, that ${\sqrt 2}f_\pi^0=130\pm
5\,MeV$ \cite{PDG}.}
  and
$\Delta^\pi_F(q^2_1)=1/(m^2_\pi+q^2_1)$ is the pion propagator.

The matrix element of the axial current ${\hat J}^a_{A,\,\mu}$ should satisfy
the partial conservation of the axial current (PCAC)
\be
{\bar u}(p')q_{1\,\mu}{\hat J}^a_{A,\,\mu}u(p)\,=\,
{\bar u}(p')\left[2M g_AF_A(q^2_1)-\frac{g_P(q^2_1)}{m_l}q^2_1\right]
\g5\frac{\tau^a}{2}u(p)\,=\,i f_\pi m^2_\pi\Delta^\pi_F(q^2_1)\,M^a_\pi
\,,  \label{PCAC}
\ee
where $M^a_\pi$ is the pion absorption amplitude. We also put
\be
g_A(q^2_1)\,=\,g_A F_A(q^2_1)\,,\quad g_A\,\equiv\,g_A(0)\,
=\,-1.2695\pm 0.0029\,
. \label{gA}
\ee
The value of the constant $g_A$ is taken from Ref.\,\cite{PDG}.
In order to fulfil Eq.\,(\ref{PCAC}) one subtracts from the induced pseudoscalar form
factor a piece
\bea
\Delta g_P(q^2_1)\,&=&\,-2g_{\pi NN} f_\pi m_l\frac{1}{q^2_1}\left[1
+\frac{Mg_A}{g_{\pi NN} f_\pi}F_A(q^2_1)\right]  \nonumber \\
\,&\approx &\,
2Mg_A m_l\frac{1}{q^2_1}\left[1-F_A(q^2_1)\right]\,\approx\,
\frac{1}{3}Mg_A m_lr^2_A\,.  \label{dgP}
\eea
Here $r_A$ is a nucleon axial radius measured independently in the
quasi--elastic neutrino scattering, $r^2_A=0.42\pm 0.04\, fm^2$
\cite{LAA}, and in the charged pion electroproduction,
$r^2_A=0.403\pm 0.030\, fm^2$ \cite{ALI}.

In deriving Eq.\,(\ref{dgP}), we use the  Goldberger--Treiman relation
\be
M|g_A|\,=\,g_{\pi NN}(0) f_\pi\,, \label{GTR}
\ee
as being satisfied exactly
and we assume a weak dependence of the  couplings
on the momentum transfer. Using the constants $g_A$, $f_{\pi^+}$
and $g_{\pi NN}$ as given  above
and the mean nucleon mass $M=938.92\,MeV$, one observes that the left-
and right hand sides of Eq.\,(\ref{GTR}) differ by 1.1\%.

The prediction for the form factor $g_P$, following from the above discussed
material, is given for the ordinary muon capture (OMC) in the hydrogen,
\be
\mu^- + p \longrightarrow \nu_\mu + n\,, \label{OMCp}
\ee
as \cite{TK}
\be
g_P(q^{\,2}_1=0.877 m^2_\mu)\, =\, \frac{2M m_\mu}{0.877 m^2_\mu
+ m^2_\pi}\,g_A\, =\, 6.87\,g_A\,=\,-8.72, \label{gOMCp}
\ee
whereas the correction, demanded by the PCAC is \cite{TK}
\be
\Delta g_P\,=\,0.34 g_A\,=\,-0.43\,.  \label{dgPOMCH}
\ee
This correction is obtained by using the dipole form factor for
$F_A(q^2_1)$ and the axial mass $m_A=1.077\,GeV$, extracted from the data
in Ref.\,\cite{ALI}, which is equivalent to taking the
nucleon axial radius $<r^2_A>^{1/2}=0.635\,fm$ \cite{ALI}.
Then the resulting value of the induced pseudoscalar constant,
as predicted by the PCAC, is
\be
g_P^{PCAC}\,=\,6.44 g_A\,=\,-8.29\,.  \label{gPPCAC}
\ee
Let us note that this value of $g_P$, obtained from the PCAC constraint,
is of fundamental importance. It is confirmed by the calculations,
performed within the framework of the effective field theory \cite{BKM,BFHM,FLMS}
that incorporates the chiral symmetry of the quantum chromodynamics.

The influence of the induced pseudoscalar form factor $g_P$ on observables
and the related extraction of this quantity from experiments
was  studied intensively in the OMC and radiative muon capture (RMC) in  nuclei and
in the electroproduction of charged soft pions. This activity has recently been
reviewed in Refs.\, \cite{MEA,BEM,TCT,GF}.

Let us note here first that in our opinion, the electroproduction of charged soft pions
cannot provide any information on the induced pseudoscalar form factor $g_P$.
Attempts to study it \cite{ BEM,CH} in the reaction
\be
e\,+\,p\,\longrightarrow\,e'\,+\,\pi^+\,+\,n\,,   \label{CPP}
\ee
at the threshold stem from the soft pion electroproduction amplitude
derived from the low energy theorem. As usually derived, this amplitude
contains the current--current amplitude and the
weak axial nucleon current. In the next step, one calculates the contributions
to the current--current amplitude. If one restricts oneself only
to the contributions  due to the nucleon Born terms, one really has an
electroproduction amplitude containing the $g_P$ form factor.
However, as shown in detail \cite{ET}, the correct calculation of the contribution to
the current--current amplitude due to the pion and the heavy axial meson emitted
in the t--channel, leads to a complete cancellation  of the weak axial nucleon
current in the electroproduction amplitude. Simultaneously,
a contact term, containing $g_A$ form factor, and the pion pole production
amplitude appear, as expected from physical intuition.

Since the extraction of the information on $g_P$ from the data on the OMC
and RMC in nuclei is extensively reviewed \cite{MEA,BEM,TCT,GF},
we restrict ourselves only to necessary comments on the RMC in
hydrogen, allowing us to proceed to the material that is the subject of this review.

For the OMC in the hydrogen, the induced pseudoscalar form factor $g_P$ only
for one value of the momentum transfer needs to be considered.
In contrast to it, the RMC in  hydrogen
\be
\mu^-\, +\, p\, \longrightarrow \,\nu_\mu\, +\,\gamma\,+\, n\,,   \label{RMCp}
\ee
allows one to study the momentum dependence of $g_P$ in a certain interval
of the values of the momentum transfer.

The RMC amplitude  generally consists of two parts. One part is due to the lepton
radiation, another one is due to the hadron radiation. In the lepton radiation amplitude,
the weak form factors depend on the four--momentum transfer
$q^L=p_1-p'_1=\nu+k-\mu=-q_1$
\footnote{The notations are obvious: $\nu$, $\mu$, $k$ is the four--momentum of the
neutrino, muon and photon, respectively.}, whereas the momentum dependence of the
weak form factors entering the hadron radiation amplitude is given by
$q^N=\nu-\mu=q^L-k$. At the end of the photon spectrum, we have $(q^L)^2\approx +m^2_\mu$
and $(q^N)^2\approx -m^2_\mu$. Thus one obtains the value of
$g_P$ in the hadron radiation amplitude larger by a factor of about 3  in comparison
with the value of $g_P$ in the lepton radiation amplitude. This enhancement factor makes
the reaction (\ref{RMCp}) attractive for the study of the sensitivity of the
high energy tail of the photon spectrum to the form factor $g_P$.

The photon spectrum was measured in the reaction (\ref{RMCp}) in a TRIUMF
experiment \cite{TRIUMF1,TRIUMF2}. The comparison of the measured spectrum with the
calculations \cite{F,BF1} provided the value of the pseudoscalar coupling constant
that is by about 50\% smaller than its value (\ref{gPPCAC}) predicted by the PCAC.
In these calculations, a relativistic RMC amplitude was used, derived
from  Feynman tree graphs. This amplitude includes also the contribution from
the $\Delta(1232)$ excitation process. Besides, it satisfies approximately the
Ward--Takahashi identities, generally derived in Ref.\,\cite{ChS}.
Subsequent studies, aiming to find an additional enhancement mechanism
of the tail of the photon spectrum, were performed  by several authors
\cite{TK,BHM,MMK,SADM,AMK,AMK1}. Only the model taking into account the
off--shell degrees of freedom of the $\Delta(1232)$ isobar is able to provide
enough enhancement of the photon spectra \cite{TK}.

One of the possible contributions to the enhancement mechanism has  been studied in detail in
Refs.\,\cite{TK,STK,TSK}. This is the contribution of the processes described by anomalous
Lagrangians. We next provide a detailed account of the construction
of such a Lagrangian for the $\pi$-$\rho$-$\omega$-$a_1$ system,
then we present the structure of the RMC amplitude arising from it. We show that
this part of the full RMC amplitude satisfies the PCAC constraint by itself.

\section{Anomalous Lagrangian  of the $\pi$-$\rho$-$\omega$-$a_1$ system }
\label{CH1}

As in the OMC, in order to calculate the capture rate,
one needs to know the effective Hamiltonian. The velocity
independent part of it is \cite{TK,RoT}
\bea
H^{\,(0)}_{\,eff}\,&=\,&\frac{1}{\sqrt{2}m_\mu}\,(1-\sl \cdot \nh )
\left[g_1\left(\sl \cdot \ve \right) + g_2 \left(\s \cdot \ve \right) +
g_3 i \left(\s \cdot \ve \times \sl \right) \right. \nonumber \\
& & \left. + g'_4 \left(\sl \cdot \ve \right) \left(\s \cdot \kh \right) +
g''_4 \left(\sl \cdot \ve \right) \left(\s \cdot \nh\right) +
g'_5 \left(\sl \cdot \kh\right) \left(\ve \cdot \nh \right) +
g'_6 \left(\ve \cdot \nh\right) \right. \nonumber \\
& & \left. + g'_7 i \left(\s \cdot \kh \times \ve\right)
+ g''_7 i \left(\s \cdot \nh \times \ve\right)
+ g'_8 \left(\sl \cdot \kh\right)\left(\s \cdot \ve \right)
+ g''_8 \left(\sl \cdot \nh\right)\left(\s \cdot \ve \right) \right. \nonumber \\
& & \left. + g'_9 \left(\sl \cdot \s\right)
+ g'_{10}\left(\s \cdot \kh\right) \left(\ve \cdot \nh \right)
+ g''_{10}\left(\s \cdot \nh\right) \left(\ve \cdot \nh \right) \right. \nonumber \\
& & \left. +g'_{11}\left(\sl \cdot \kh\right)\left(\s \cdot \kh\right)
\left(\ve \cdot \nh \right)
+g''_{11}\left(\sl \cdot \kh\right)\left(\s \cdot \nh\right)
\left(\ve \cdot \nh \right) \right]\,.  \label{H0eff}
\eea
Here $\vec {\sigma}_l$ ($\vec \sigma$) are the lepton (nucleon) spin Pauli matrices,
$\hat \nu$ ($\hat k$) is the unit vector in the direction of the neutrino (photon)
momentum vector $\vec \nu$ ($\vec k$) and $\ve_\lambda$ is the photon polarization.
The most important form factors are $g_1$, $g_2$ and $g_3$. All other $g_i$ contain
at least one damping factor $1/M$.

The aim of any model Lagrangian  of the interaction of a hadron system with an
external electroweak fields is to provide the form factors $g_i$ entering
Eq.\,(\ref{H0eff}). Usually, one considers processes, described by the
normal Lagrangians \cite{BHM,BF1,MMK,SADM,STKN}, where a natural parity
of the in- and outcoming channels does not change. The natural parity of a
particle is defined for bosons only and it is
$P_n=P\,(-1)^J$, where $P$ is the intrinsic parity and $J$ is the spin of
the particle. The natural parity of the channel is defined as the product
of the natural parities of the channel particles.

The RMC amplitude, presented in Ref.\,\cite{STKN}, is derived from a non--anomalous
Lagrangian of the $\pi\rho\omega a_1$ system that reflects the
$SU(2)_L\times SU(2)_R$ hidden local symmetry \cite{BKY,M,KM,STG}.
This amplitude extends the amplitude obtained from the low energy theorem
to higher values of the photon and weak current momenta.

Here our goal is to construct the RMC  amplitude which contributes to the
anomalous processes. Let first discuss the generalities related
to the construction of the necessary anomalous Lagrangian.

In  meson physics,
anomalous processes are defined as processes in which the natural parity
is not preserved. The value of $P_n$ for some bosons is given in table 1.

\vskip 0.2cm

\begin{center}
Table 1. Natural parity of some bosons.

\vskip 0.1cm

\begin{tabular}{|l||c|c|c|}\hline 
boson   &    $P$ & $J$ & $P_n$  \\\hline
$\sigma$&    +1  &  0  &  +1  \\
$\pi$   &    -1  &  0  &  -1  \\
$\rho$  &    -1  &  1  &  +1  \\
$\omega$&    -1  &  1  &  +1  \\
$a_1$   &    +1  &  1  &  -1  \\
$f_1(1285)$& +1  &  1  &  -1  \\
$\gamma$&    -1  &  1  &  +1  \\\hline
\end{tabular}
\end{center}

\vskip 0.2cm

Then the vertices  $\rho\rightarrow \pi\pi$ and $a_1\rightarrow \rho\pi$
are described by the non--anomalous Lagrangian, whereas the vertices
$\rho\rightarrow\gamma\pi$  and $\omega\rightarrow\gamma\pi$ are anomalous
and play an important role in describing the deuteron electromagnetic
form factors. As we shall see soon, the anomalous vertices
$a_1\rightarrow\rho\omega$ and $\rho\rightarrow\omega\pi$ enter into
the amplitudes for the process (\ref{RMCp}).

There exists a unique way to construct the interaction Lagrangian that
would violate the natural parity and would simultaneously conserve the
intrinsic parity and would be Lorentz invariant:  the
Levi--Civita pseudotensor should be used. So anomalous processes
are defined as
processes, described by a Lagrangian containing the pseudotensor
$\vep_{\alpha\beta\gamma\lambda}$. Let us note that this definition of
the anomalous processes is more general than the definition using
the natural parity, as it can be extended from  pure meson processes
also to processes, where mesons interact also with the gauge
bosons of the electroweak interaction and subsequently, with fermions.

Let us note that the term "anomalous" refers to the axial Abelian as well as non--Abelian
anomaly\footnote {The axial--vector anomaly is frequently called the chiral anomaly
\cite{SA}.}, yielding an anomalous breaking of the chiral symmetry in the
theory of quantum fields \cite{W2}. The first such chiral anomaly was observed in the width
of the decay
\be
\pi^0\,\rightarrow\,\gamma\,\gamma\,.  \label{pi0gg}
\ee
The observed decay width turned out to be enhanced by three orders of magnitude
in comparison with the one derived from the models incorporating spontaneously
and explicitly broken chiral symmetry \cite{DGH,W2}. Anomalous behavior of the
decay width for the process (\ref{pi0gg}) indicates the existence of a new
mechanism of the chiral symmetry breaking. It was found \cite{BJ,SAD} that it is the
regularization of the  Feynman one--loop diagrams that produces the chiral symmetry
breaking.

An elegant study of the chiral anomaly  has later been performed
within the path integral method in Ref.\,\cite{FU}. Non--invariance under
the chiral transformations of the
measure in the path integral over the fermion  fields
means that the functional integral is not invariant  and the chiral symmetry
is broken. This formally provides Jacobi determinant different from  identity.
It follows that the logarithm of such a determinant leads to an anomaly function
\cite{DGH,W2} that can be given in  terms of the gauge fields of the
electroweak interaction, but it does not depend on the
quark fields. Finally, its presence produces in the effective Lagrangian
the term
\be
{\cal L}_{\pi^0\gamma\gamma}\,=\,-\frac{e^2}{32 \pi^2 f_\pi}\,
\vep_{\alpha\beta\gamma\lambda}F_{\alpha\beta}(x)\,F_{\gamma\lambda}(x)\,
\pi^0(x)\,,  \quad F_{\alpha\beta}(x)\,=\,\partial_\alpha A_\beta(x)
\,-\,\partial_\beta A_\alpha(x)\,.  \label{Lpi0gg}
\ee
This Lagrangian can be qualified as anomalous in accord with both definitions
of anomalous Lagrangians discussed above and it provides the
correct value of the decay width for the process (\ref{pi0gg}).

It is important to note that the form of the anomaly function is rather independent
of the detailed features of the theory \cite{W2}. However, it should
satisfy the Wess--Zumino consistency conditions \cite{WZ} that depend only
on the properties of the anomalously broken symmetry group. These conditions
strongly restrict the form of the anomaly function, but they do not
fix it uniquely.

The independence of the anomaly function on the details of the theory is
useful if one would like to pass from the quantum chromodynamics (QCD) to the low energy
effective field theories given in terms of the Goldstone bosons  and
the gauge fields of the electroweak interaction. The consistency conditions
will be the same as in QCD and it can  be shown that the anomaly function
will be the same, too \cite{GH,W2}.

At the level of effective low energy Lagrangians, the term,
breaking the chiral symmetry, should be present in  the Lagrangian
from the very beginning and it is called the anomalous Lagrangian. It
follows from what we said above that its change under the
infinitesimal chiral transformation is fixed  by the anomaly
function.

The form of the anomalous Lagrangian for the Goldstone bosons of the
spontaneously broken global chiral group $SU(3)_L\times SU(3)_R$ was
derived in Ref.\,\cite{WZ}. However, this Lagrangian cannot be given
in a simple closed form \cite{DGH,W2}, but it is possible for the
anomalous action \cite{WI}. Without gauge fields, this action is invariant
under the global transformations of the chiral group
\be
        G_g\,=\,\left[SU(3)_L\times SU(3)_R\right]_g\,,   \label{Gg}
\ee
whereas the related anomalous Lagrangian is not invariant.
It is possible to extend this global symmetry to the local one by introducing
the gauge fields of the electroweak interaction into the anomalous action,
the fields of the photon and of the intermediate bosons $W^{\pm}$ and $Z^0$ \cite{WI}.
Then the gauged Wess--Zumino--Witten action will contain terms, describing
anomalous processes taking place between the $\pi$-, $K$- and $\eta$
mesons in the presence of the external electroweak fields, including
the Lagrangian (\ref{Lpi0gg}), responsible for the decay (\ref{pi0gg}).

Anomalous processes are important not only in the case of the Goldstone bosons,
but are abundantly observed in the reactions with vector mesons, such as $\omega$
and $\rho$ mesons, e.g. the radiative decay of these mesons,
\be
B\,\rightarrow\,\pi\,+\,\gamma\,, \quad B\,=\,\omega,\,\rho\,,\label{RDB}
\ee
is the anomalous process, as it follows from table 1.

For an extension of the theory to incorporate the vector mesons, it is important to
recognize that any such  theory should be free of axial  anomaly,
as it follows from the gauge invariance \cite{W2}.
In other words, the anomaly functions, arising in different sectors of the theory,
should compensate among themselves. It turns out that this restriction
does not allow one to incorporate the vector mesons into the anomalous Lagrangians
as Yang--Mills gauge fields. This follows from the fact that in this case,
the anomaly function, arising in the vector meson sector of the anomalous Lagrangian,
differs from the
one, derived from the QCD. Then such a theory is in contradiction with the
fundamental postulate \cite{WPH} that  any effective low energy theory of hadrons
should strictly respect  symmetries imbedded in the QCD and the consequences
following from it, including the mode in which these symmetries are broken.
Since the anomaly function is a direct consequence of the anomalously broken
chiral symmetry of the QCD, the same anomaly function should be immanent also
to any anomalous hadron Lagrangian.

On the contrary, introducing the vector mesons within the framework of the hidden
local symmetries (HLS) provides a theory of the interaction of the mesons
with the electroweak fields that is free of the above mentioned
defects \cite{BKY,M,KM,FKTUY}. Methodologically, the construction of the anomalous
HLS Lagrangian does not differ from that of the non--anomalous one. The extension
of the symmetry (\ref{Gg}) by the group
$G_l\equiv \left[SU(3)_L\times SU(3)_R\right]_l$ of the local transformations
needs introduction both of the related massless gauge fields and non--physical compensators.
In the next step, after the spontaneously breaking of the extended symmetry,
the compensator fields disappear, whereas the gauge fields acquire the mass.
These new gauge fields are identified with the physical vector and axial--vector
mesons. Then the procedure results in the appearance of the
anomalous invariants, consisting of the fields of particles, which anomalous
processes one would like to describe of: of the Goldstone-, vector- and
axial--vector meson fields and of the fields of the electroweak interaction.
In general, the anomalous Lagrangian can be chosen as the sum of the
Wess--Zumino--Witten Lagrangian and of the linear combination of the anomalous invariants
\cite{BKY,M,KM}. Thus, such a Lagrangian contains free parameters that can be
fixed by analyzing anomalous processes, or by imposing additional
semi--phenomenological conditions.

The anomalous Lagrangian, containing both the vector-  ($\rho,\,\omega$) and axial--vector
($a_1,\,f_1$) mesons, was first constructed in Ref.\,\cite{KM}. Such a Lagrangian
contains 14 invariants. In Ref.\,\cite{KM}, only the electromagnetic anomalous
processes were studied. The extension of this model to the weak anomalous
processes was accomplished in Refs.\,\cite{STK,TSK}. To carry out this step, it is advantageous
to use the anomalous invariants, constructed in Ref.\,\cite{KM} and
to express the external gauge fields using both the photon field and the
boson fields $W^{\pm}$ and $Z^0$.

The most general anomalous action of the $\pi\rho\omega a_1 f_1$ system reads \cite{KM}
\bea
\Gamma_{an}[\xi_L,\xi_R,\xi_M,L,R,{\cal L},{\cal R}]\,& = &\,
\Gamma^{cov}_{WZW}[U,{\cal L},{\cal R}] \nonumber  \\
& & \,+\sum_{i=1}^{14}\,\int_{M^4}\,c_i\,
{\cal L}_i[\xi_L,\xi_R,\xi_M,L,R,{\cal L},{\cal R}]\,. \label{GAN}
\eea
Here $\Gamma^{cov}_{WZW}[U,{\cal L},{\cal R}]$ is the covariant
Wess-Zumino-Witten action containing pseudoscalars and the electroweak
fields. It already satisfies the anomaly constraints. Generally, the
14 independent (homogenous)
terms in the r.\,h.\,s.\, of Eq.\,(\ref{GAN}) are given in Eqs.\,(3.8) and
(3.9) of Ref.\,\cite{KM}.
As the terms ${\cal L}_1$-${\cal L}_8$ contain at least 4 particles in
each vertex, only the terms ${\cal L}_9$-${\cal L}_{14}$ are of interest
for our purpose.

The covariant Wess-Zumino-Witten anomalous action of pseudoscalars
\newline reads \cite{BKY,M,WI}
\be
\Gamma^{cov}_{WZW}[U,{\cal L},{\cal R}]\,=\,-i\frac{N_c}{240\,\pi^2}\,
\int_{M^5}\,{\it Tr}[\al^5]_{covariantized}\,,  \label{Gcov}
\ee
where $N_c$ is the number of colors and $\al$ is a differential
one-form
\be
\al\,=\,(\partial_\mu U) U^{\dagger} dx_\mu\,,\qquad\,
U(x)\,=\,exp[-i\Pi^a(x)\tau^a/f_\pi]\,\equiv\,\xi^2\,,
\ee
and ${\cal L}_\mu,{\cal R}_\mu$ are the external gauge fields.

The contribution from the action (\ref{Gcov}) to the 3-point Lagrangian
of interest is \cite{STK}
\be
{\cal L}_{WZW}\,=\,i\frac{e^2}{8\pi^2 f_\pi}\, \vep_{\kappa \lambda
 \mu \nu}
(\partial_\kappa \widetilde {\cal B}_\lambda)(\partial_\mu \vec {\cal V}_\nu
\,\cdot\,\vec{\Pi})\,,   \label{LWZW}
\ee
where ${\widetilde {\cal B}}_\lambda$ is an electroweak neutral field and
$\vec {\cal V}_\nu$ is a weak vector field.

In the homogenous terms, we include both the electromagnetic and weak interactions,
but we omit the field of the $f_1$ meson. Keeping only the 3--particle terms,
the anomalous Lagrangian of the $\pi\rho\omega a_1$ system \cite{STK,TSK}
is obtained
\be
\bar {{\cal L}}_{an}\,=\,\sum_{i=7}^{10}\,\bar{c}_i\,\bar {{\cal L}}_i\,, \label{Lan}
\ee
where  the $\bar {{\cal L}}_i$ terms are \footnote{In the proceedings version
of the Ref.\,\cite{STK}, the factor 2g  is lacking
at the r.h.s. of Eqs.\,(50) and (51), whereas the factor g in the first
term in the braces is superfluous; the same is true for Eqs.\,(52) and (53),
but with $g\rightarrow e$.}
\bea
\bar {\cal L}_7\,&=&\,2ig_\rho\vep_{\kappa \lambda \mu \nu}\,\{
\partial_\kappa \omega_\lambda\,[( g_\rho \vec {\rho}_\mu\,-\,e\vec {\cal V}_\mu)
\cdot
(\frac{1}{f_\pi}\,\partial_\nu\vec {\pi}\,+\,e\vec {\cal A}_\nu)]
\nonumber  \\
& & +\,(g_\rho\omega_\kappa\,-\,\frac{1}{3}e\,{\cal B}_\kappa)\,
[(\partial_\lambda \vec {\rho}_\mu) \cdot (\frac{1}{f_\pi}\,
\partial_\nu \vec {\pi}\,+\,e\vec {\cal A}_\nu)]\}\,,  \label{LB73}  \\
\bar {\cal L}_8\,&=&\,-2ig_\rho\vep_{\kappa \lambda \mu \nu}\,\{
\partial_\kappa \omega_\lambda\,[( g_\rho \vec {\rho}_\mu\,-\,e\vec {\cal V}_\mu)
\cdot
(g_\rho \vec {a}_\nu\,+\,\frac{1}{2f_\pi} \partial_\nu \vec {\pi})]
\nonumber  \\
& & +\,(g_\rho\omega_\kappa\,-\,\frac{1}{3}e\,{\cal B}_\kappa)\,
[(\partial_\lambda \vec {\rho}_\mu) \cdot (g_\rho \vec {a}_\nu\,+\,
\frac{1}{2f_\pi}\,\partial_\nu \vec {\pi})]\}\,,  \label{LB83}   \\
\bar {\cal L}_9\,&=&\,2ie\vep_{\kappa \lambda \mu \nu}\,\{
\frac{1}{3}\partial_\kappa {\cal B}_\lambda\,
[(g_\rho \vec {\rho}_\mu\,-\,e\vec {\cal V}_\mu) \cdot
(\frac{1}{f_\pi}\,\partial_\nu \vec {\pi}\,+\,e\vec {\cal A}_\nu)] \nonumber  \\
& & +\,(g_\rho\omega_\kappa\,-\,\frac{1}{3}e\,{\cal B}_\kappa)\,
[(\partial_\lambda \vec {\cal V}_\mu)(\frac{1}{f_\pi}\,\partial_\nu \vec {\pi}
\,+\,e\vec {\cal A}_\nu)]\}\,,  \label{LB93}   \\
\bar {{\cal L}}_{10}\,&=&\,-2ie\vep_{\kappa \lambda \mu \nu}\,\{
\frac{1}{3}\partial_\kappa {\cal B}_\lambda\,
[(g_\rho \vec {\rho}_\mu\,-\,e \vec {\cal V}_\mu) \cdot
(g_\rho \vec {a}_\nu\,+\,\frac{1}{2f_\pi} \partial_\nu \vec {\pi})]
\nonumber  \\
& & +\,(g_\rho\omega_\kappa\,-\,\frac{1}{3}e\,{\cal B}_\kappa)\,
[(\partial_\lambda \vec {\cal V}_\mu)(g_\rho \vec {a}_\nu\,
+\,\frac{1}{2f_\pi} \partial_\nu \vec {\pi})]\}\,.  \label{LB103}
\eea
 External fields $\vec {\cal V}$ and $\vec {\cal A}$
correspond to the gauge fields of \mbox{the Standard Model
\cite{BKY}}
\bea
{\cal V}^{\pm}_\mu\,&=&\,-{\cal
A}^{\pm}_\mu\,=\,\frac{1}{\sin\,\Theta_w}\,
{\cal W}^{\pm}_\mu\,\cos\,\Theta_c\,,  \label{VPM} \\
{\cal V}^3_\mu\,&=&\,{\widetilde{\cal B}}_\mu\,+\,\cot\,(2\,\Theta_w)\,{\cal
Z}_\mu\,=\, {\cal B}_\mu\,+\,\frac{1}{\sin\,(2\,\Theta_w)}\,{\cal Z}_\mu\,,
\label{V3}  \\
{\cal A}^3_\mu\,&=&\,-\frac{1}{\sin\,(2\,\Theta_w)}\,{\cal
Z}_\mu\,. \label{A3}
\eea
The constants $\bar{c}_i$  are
\bea
\bar{c}_7\,&=&\,\tilde{c}_7\,+\,\frac{1}{2}\tilde{c}_8\,=\,c_9\,,\quad
\bar{c}_8\,=\,\tilde{c}_8\,=\,c_9\,-\,2c_{10}\,,  \nonumber \\
\bar{c}_9\,&=&\,\tilde{c}_9\,+\,\frac{1}{2}\tilde{c}_{10}
\,=\,c_{12}\,,  \quad
\bar{c}_{10}\,=\,\tilde{c}_{10}\,=\,c_{12}\,-\,2c_{13}\,. \label{bci}
\eea
The constants ${\tilde c}_i$ were first determined in
Ref.\,\cite{KM}. The progress in acquiring the data on  several reactions \cite{PDG98}
allowed in Ref.\, \cite{TSK} to improve the analysis considerably. The difference
between the new data \cite{PDG} and the data \cite{PDG98} is not so dramatic and only
the constants ${\tilde c}_7$ and ${\tilde c}_9$ slightly changed in comparison with \cite{TSK}
\be
\tilde{c}_7=8.72\times 10^{-3}\,,\,\, \tilde{c}_8=-1.07\times 10^{-1}\,,
\,\, \tilde{c}_9=9.76\times 10^{-3}\,,\,\, \tilde{c}_{10}=7.59\times
10^{-2}\,.  \label{ctis}
\ee
Let us note that we prefer to choose $\tilde{c}_9$ as in \cite{TSK},
by averaging the data on the processes $\rho^\pm \rightarrow \pi^\pm\gamma$
and $\omega\rightarrow\pi^0\gamma$, since the data on the reaction
$\rho^0\rightarrow\pi^0\gamma$ show clear tendency to move to the data
on the charged $\rho$ meson radiative decay \cite{PDG}.

Having the homogenous part of the anomalous Lagrangian at our disposal,
we pass to the construction of the anomalous RMC amplitude.

\section{Structure of the anomalous RMC amplitude}
\label{CH2}

The contribution to the RMC amplitude due to the Wess--Zumino term
(\ref{LWZW}) is presented in Fig.\,1. It reads
\be
J^{a,WZ}_{\mu\nu}=-\frac{g_{\pi NN}}{8\pi^2 f_\pi}\,\vep_{\mu\nu\eta\alpha}\,
k_\eta q_\alpha\,\Delta^\pi_F(q^2_1)\,\Gamma^a_5\,,
\label{WZC}
\ee
where
\be
\Gamma^a_5\,=\,{\bar u}(p')\g5\tau^a u(p)\,.  \label{Ga5}
\ee
This amplitude satisfies the continuity equations
\be
k_\nu J^{a,WZ}_{\mu\nu}\,=\,q_\mu J^{a,WZ}_{\mu\nu}\,=\,0\,.  \label{CCJWZ}
\ee
Since this amplitude contains the pion propagator, it contributes to the
form factor $g_P$. However, $q_1=q-k=\mu-\nu-k=-q^L$ and this contribution
has no enhancement factor. The influence of this amplitude on the photon spectrum was
earlier calculated in Ref.\,\cite{FLMS1} and it was found to be negligible.\\
\vskip 0.35cm
\begin{center}
  \includegraphics[height=0.25\textheight]{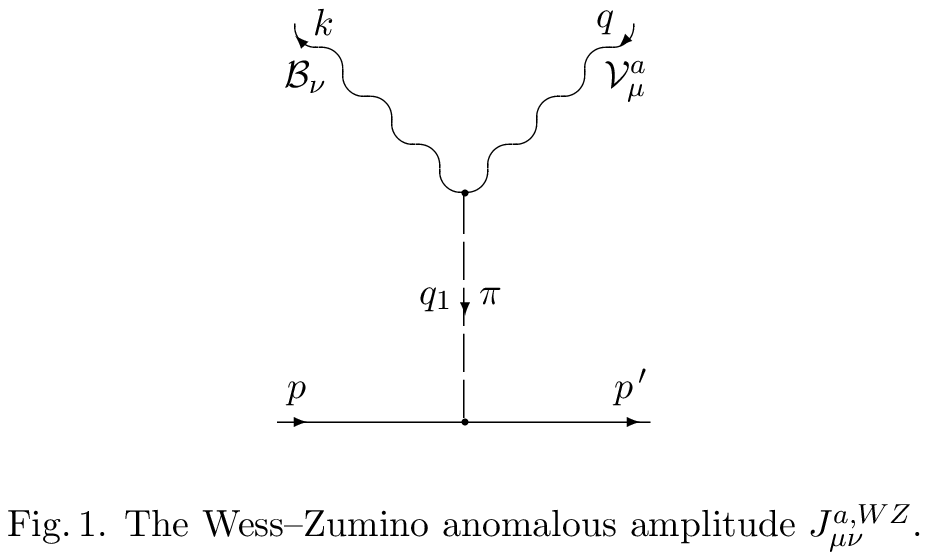}
\end{center}
\vskip .35cm

Next the contribution from the Lagrangian $\bar {\cal L}_7$ is presented in more detail.
It contains eight terms
\bea
\bar {\cal L}_7\,&=&\,\sum_{i=1}^8 \bar {\cal L}_{7,i} \nonumber \\
&=&\,2ig_\rho\vep_{\mu\nu\kappa\lambda}\left\{\,
(\pa_\mu \omega_\nu)\left[\frac{g_\rho}{f_\pi}(\vec \rho_\kappa\cdot \pa_\lambda \vec \pi)
-\frac{e}{f_\pi}({\vec {\cal V}}_\kappa\cdot\pa_\lambda\vec \pi)
+eg_\rho (\vec \rho_\kappa\cdot {\vec {\cal A}}_\lambda)
\right.\right.\nonumber \\
&&\left.\left.\,-e^2 ({\vec {\cal V}}_\kappa \cdot{\vec{\cal A}}_\lambda)\right]
+\frac{1}{f_\pi}\,\left[g_\rho \omega_\mu-\frac{e}{3}{\cal B}_\mu\right]
(\pa_\nu{\vec\rho}_\kappa\cdot\pa_\lambda{\vec\pi})
\right.\nonumber \\
&&\left.\,+e\,\left[g_\rho \omega_\mu-\frac{e}{3}{\cal B}_\mu\right]
(\pa_\nu{\vec\rho}_\kappa\cdot{\vec {\cal A}}_\lambda)\,\right\}\,. \label{LSBI}
\eea
Using particular terms, the contributions to the anomalous
RMC amplitude are calculated. From the vertex $\bar {\cal L}_{7,1}$, one generates 3
Feynman amplitudes, presented in Fig.\,2a, Fig.\,2b and Fig.\,2c. They are of the form
\bea
J^{a}_{\mu\nu}(7,1a)&=&-2m^2_\omega m^2_\rho\frac{g_\rho g_{\pi NN}}{g_\omega f_\pi}
\vep_{\alpha\beta\lambda\kappa}k_\kappa
q_{1\lambda}\Delta^\pi_F(q^2_1)\Delta^\omega_{\alpha\nu}(k)
\Delta^\rho_{\beta\mu}(q)\Gamma^a_5\,, \\ \label{J71a}
J^{a}_{\mu\nu}(7,1b)&=&-im^2_\omega\frac{g^3_\rho}{g_\omega}
\vep_{\alpha\beta\lambda\kappa}\,k_\kappa q_\lambda  q_\mu
\Delta^\pi_F(q^2)\Delta^\omega_{\alpha\nu}(k)\Delta^\rho_{\beta\eta}(q_1)
\Gamma^a_\eta  \nonumber \\
&=&-i f_\pi q_\mu \Delta^\pi_F(q^2) M^{a}_\nu(7,1c)\,, \\ \label{J71b}
M^{a}_\nu(7,1c)&=&m^2_\omega\frac{g^3_\rho}{g_\omega f_\pi}
\vep_{\alpha\beta\lambda\kappa}\,k_\kappa q_\lambda
\Delta^\omega_{\alpha\nu}(k)\Delta^\rho_{\beta\eta}(q_1)
\Gamma^a_\eta\,.   \label{M71c}
\eea
Here
\be
\Gamma^a_\eta={\bar u}(p')(\gamma_\eta
-\frac{\kappa_V}{2M}\sigma_{\eta\delta}q_{1\delta})u(p) \,. \label{Gae}
\ee
Both amplitudes, $J^{a}_{\mu\nu}(7,1a)$ and $J^{a}_{\mu\nu}(7,1b)$, contain the pion propagator
and contribute to $g_P$. However, only  the  pion propagator of the amplitude $J^{a}_{\mu\nu}(7,1b)$
provides the enhancement factor $\approx\,3$.

Calculating the divergence yields
\bea
k_\nu J^{a}_{\mu\nu}(7,1a)\,&=&\,k_\nu J^{a}_{\mu\nu}(7,1b)\,=\,0\,, \\ \label{kJ71ab}
q_\mu J^{a}_{\mu\nu}(7,1a)\,&=&\,-2m^2_\omega\vep_{\alpha\beta\lambda\kappa}k_\kappa
q_\beta q_{1\lambda}\Delta^\pi_F(q^2_1)\Delta^\omega_{\alpha\nu}(k)
\Gamma^a_5\,,  \\  \label{qJ71a}
q_\mu J^{a}_{\mu\nu}(7,1b)\,&=&\,-i f_\pi q^2 \Delta^\pi_F(q^2) M^{a}_\nu(7,1c)\,.
\label{qJ71b}
\eea
Relative to the index $\nu$, attached to the photon line, the amplitudes are transverse
separately.

The vertex $\bar {\cal L}_{7,2}$ generates the amplitude presented in Fig.\,2d.
Explicitly we have
\bea
J^{a}_{\mu\nu}(7,2)\,&=&\,2m^2_\omega\frac{g_\rho g_{\pi NN}}{g_\omega f_\pi}
\vep_{\alpha\mu\lambda\kappa}k_\kappa
q_{1\lambda}\Delta^\pi_F(q^2_1)\Delta^\omega_{\alpha\nu}(k)
\Gamma^a_5\,, \\  \label{J72}
k_\nu J^{a}_{\mu\nu}(7,2)\,&=&\,0\,,\quad
q_\mu J^{a}_{\mu\nu}(7,2)\,=\,-q_\mu J^{a}_{\mu\nu}(7,1a)\,. \label{kqJ72}
\eea
Again, the amplitude is transverse in the electromagnetic sector. As follows from
the second part of Eq.\,(\ref{kqJ72}), the weak vector amplitudes $J^{a}_{\mu\nu}(7,1a)$ and
$J^{a}_{\mu\nu}(7,2)$ satisfy the CVC hypothesis.

From the vertex $\bar {\cal L}_{7,3}$ we have also only one amplitude of Fig.\,2e
\bea
J^{a}_{\mu\nu}(7,3)\,&=&\,i m^2_\omega\frac{g^3_\rho}{g_\omega}
\vep_{\mu\alpha\beta\kappa}k_\kappa\Delta^\omega_{\alpha\nu}(k)
\Delta^\rho_{\beta\eta}(q_1)\Gamma^a_\eta\,,   \\  \label{Ja73}
k_\nu J^{a}_{\mu\nu}(7,3)\,&=&\,0\,,\quad
q_\mu J^{a}_{\mu\nu}(7,3)\,=\,if_\pi\,M^{a}_\nu(7,1c)\,. \label{kqJ73}
\eea
It is seen  from Eq.\,(\ref{qJ71b}) and Eq.\,(\ref{kqJ73}) that
the axial amplitudes $J^{a}_{\mu\nu}(7,1b)$ and $J^{a}_{\mu\nu}(7,3)$
satisfy the PCAC
\be
q_\mu [J^{a}_{\mu\nu}(7,1b)\,+\,J^{a}_{\mu\nu}(7,3)]\,=\,
if_\pi\,m^2_\pi \Delta^\pi_F(q^2)\,M^{a}_\nu(7,1c)\,. \label{qJ71b3}
\ee
\vskip 0.35cm
\begin{center}
  \includegraphics[height=0.7\textheight]{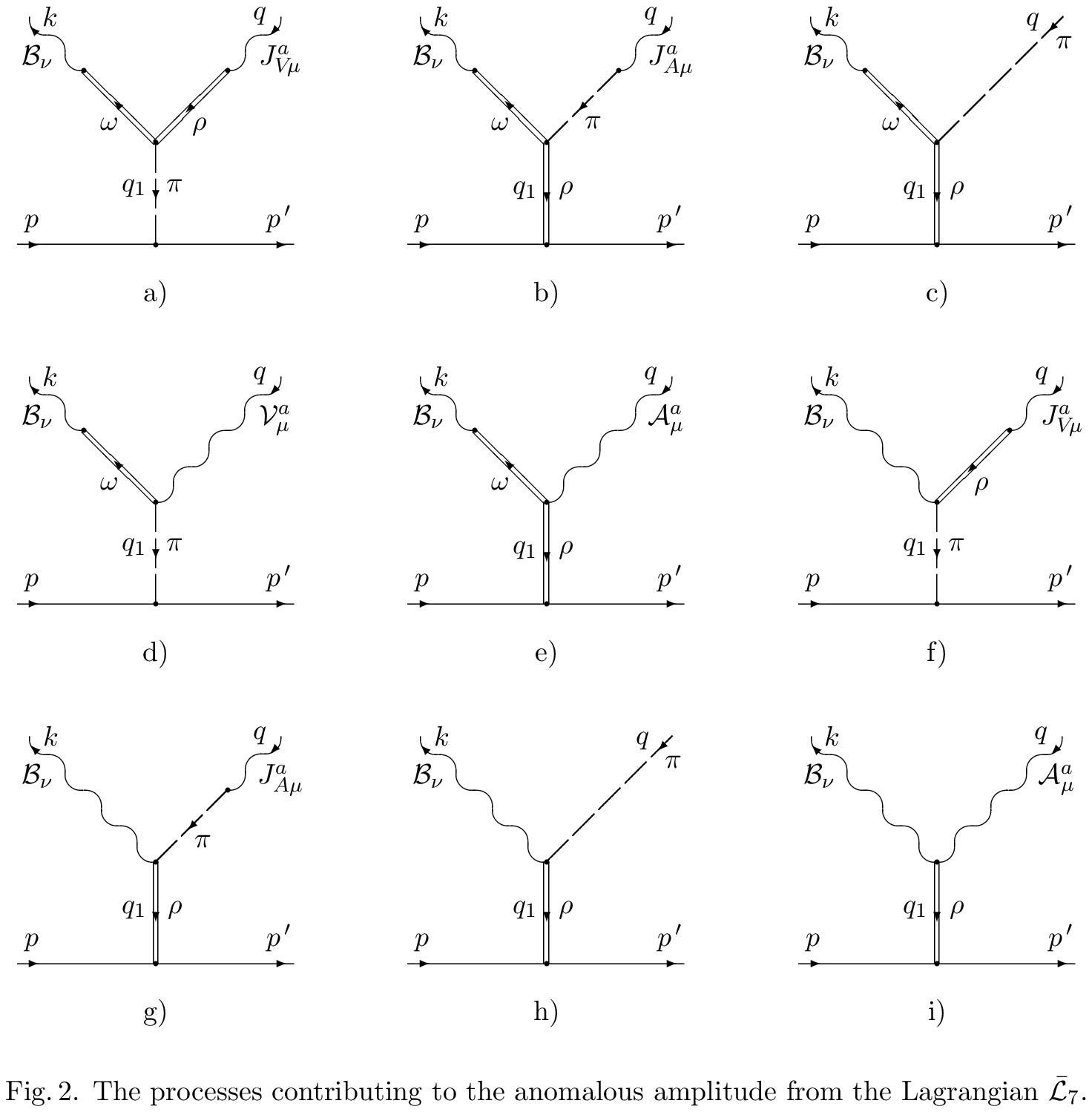}
\end{center}
\vskip 0.35cm

The vertex $\bar {\cal L}_{7,4}$ does not contribute to processes triggered
by the charged current.

In its turn, the vertex $\bar {\cal L}_{7,5}$ generates 3 amplitudes, again
represented by the graphs Fig.\,2a, Fig.\,2b and Fig.\,2c
\bea
J^{a}_{\mu\nu}(7,5a)&=&2m^2_\omega m^2_\rho\frac{g_\rho g_{\pi NN}}{g_\omega f_\pi}
\vep_{\alpha\beta\lambda\kappa}
q_\alpha q_{1\lambda}\Delta^\pi_F(q^2_1)\Delta^\omega_{\kappa\nu}(k)
\Delta^\rho_{\beta\mu}(q)\Gamma^a_5\,, \\ \label{J75a}
J^{a}_{\mu\nu}(7,5b)&=&-im^2_\omega\frac{g^3_\rho}{g_\omega}
\vep_{\alpha\beta\lambda\kappa}\,q_{1\alpha} q_\lambda  q_\mu
\Delta^\pi_F(q^2)\Delta^\omega_{\kappa\nu}(k)\Delta^\rho_{\beta\eta}(q_1)
\Gamma^a_\eta  \nonumber \\
&=&-i f_\pi q_\mu \Delta^\pi_F(q^2) M^{a}_\nu(7,5c)\,, \\ \label{J75b}
M^{a}_\nu(7,5c)&=&m^2_\omega\frac{g^3_\rho}{g_\omega f_\pi}
\vep_{\alpha\beta\lambda\kappa}\, q_\lambda q_{1\alpha}
\Delta^\omega_{\kappa\nu}(k)\Delta^\rho_{\beta\eta}(q_1)
\Gamma^a_\eta\,.   \label{M75c}
\eea
Again, both amplitudes, $J^{a}_{\mu\nu}(7,5a)$ and $J^{a}_{\mu\nu}(7,5b)$, contain the pion propagator
and contribute to $g_P$. However, only  the  pion propagator of the amplitude $J^{a}_{\mu\nu}(7,5b)$
provides the enhancement factor $\approx\,3$.
The divergence of these amplitudes reads
\be
k_\nu J^{a}_{\mu\nu}(7,5a)=2 m^2_\rho\frac{g_\rho g_{\pi NN}}{g_\omega f_\pi}
\vep_{\alpha\beta\lambda\kappa}k_\kappa
q_\alpha q_{1\lambda}\Delta^\pi_F(q^2_1)
\Delta^\rho_{\beta\mu}(q)\Gamma^a_5\,,  \label{kJ75a}
\ee
\be
q_\mu J^{a}_{\mu\nu}(7,5a)=0\,,  \label{qJ75a}
\ee
\be
k_\nu J^{a}_{\mu\nu}(7,5b)=-i\frac{g^3_\rho}{g_\omega}
\vep_{\alpha\beta\lambda\kappa}\,k_\kappa q_{1\alpha} q_\lambda  q_\mu
\Delta^\pi_F(q^2)\Delta^\rho_{\beta\eta}(q_1)\Gamma^a_\eta  \,,  \label{kJ75b}
\ee
\be
q_\mu J^{a}_{\mu\nu}(7,5b)=-i f_\pi q^2 \Delta^\pi_F(q^2) M^{a}_\nu(7,5c)\,. \label{qJ75b}
\ee
Now the amplitudes are not separately transverse in the electromagnetic sector.
On the contrary, the weak vector amplitude $J^{a}_{\mu\nu}(7,5a)$ satisfies the CVC by itself.

From the vertex $\bar {\cal L}_{7,6}$, one again obtains 3 amplitudes, given in Fig.\,2f,
Fig.\,2g and Fig.\,2h
\be
J^{a}_{\mu\nu}(7,6a)=2 m^2_\rho\frac{g_\rho g_{\pi NN}}{g_\omega f_\pi}
\vep_{\nu\kappa\beta\lambda}q_\kappa q_{1\lambda}\Delta^\pi_F(q^2_1)
\Delta^\rho_{\beta\mu}(q)\Gamma^a_5\,,   \label{J76a}
\ee
\bea
J^a_{\mu\nu}(7,6b)&=&-i\frac{g^3_\rho}{g_\omega}
\vep_{\nu\kappa\beta\lambda}
q_\mu q_\lambda q_{1\kappa}\Delta^\pi_F(q^2)
\Delta^\rho_{\beta\eta}\Gamma^a_\eta \nonumber \\
&=&-i f_\pi q_\mu \Delta^\pi_F(q^2) M^{a}_\nu(7,6h)\,,  \label{J76b}
\eea
\be
M^{a}_\nu(7,6h)=\frac{g^3_\rho}{g_\omega f_\pi}
\vep_{\nu\kappa\beta\lambda}q_\lambda q_{1\kappa}
\Delta^\rho_{\beta\eta}\Gamma^a_\eta\,.
\label{M76h}
\ee
In deriving these equations, we used the relation $g_\omega=3 g_\rho$.
It is the amplitude $J^a_{\mu\nu}(7,6b)$ that contains the enhancement factor.
The divergence of the amplitudes $J^a_{\mu\nu}(7,6a)$ and $J^a_{\mu\nu}(7,6b)$ is
\be
k_\nu J^{a}_{\mu\nu}(7,6a)=2 m^2_\rho\frac{g_\rho g_{\pi NN}}{g_\omega f_\pi}
\vep_{\nu\kappa\beta\lambda}k_\nu q_\kappa q_{1\lambda}\Delta^\pi_F(q^2_1)
\Delta^\rho_{\beta\mu}(q)\Gamma^a_5\,,   \label{kJ76a}
\ee
\be
q_\mu J^{a}_{\mu\nu}(7,6a)=0\,,  \label{qJ76a}
\ee
\be
k_\nu J^a_{\mu\nu}(7,6b)=i\frac{g^3_\rho}{g_\omega}
\vep_{\alpha\beta\lambda\kappa}k_\kappa
q_{1\alpha}q_\lambda q_\mu \Delta^\pi_F(q^2)
\Delta^\rho_{\beta\eta}\Gamma^a_\eta\,,   \label{kJ76b}
\ee
\be
q_\mu J^a_{\mu\nu}(7,6b)=-i f_\pi q^2 \Delta^\pi_F(q^2) M^{a}_\nu(7,6h)\,. \label{qJ76b}
\ee
It follows from Eq.\,(\ref{kJ75b}) and Eq.\,(\ref{kJ76b}) that
\be
k_\nu [J^a_{\mu\nu}(7,5b)+J^a_{\mu\nu}(7,6b)]=0\,. \label{kJ756b}
\ee

The vertex $\bar {\cal L}_{7,7}$ yields the amplitude of  Fig.\,2e.
Together with the divergence, this amplitude  is
\be
J^{a}_{\mu\nu}(7,7)=i m^2_\omega \frac{g^3_\rho}{g_\omega}
\vep_{\alpha\beta\mu\kappa}q_{1\alpha} \Delta^\omega_{\nu\kappa}(k)
\Delta^\rho_{\beta\eta}(q_1)\Gamma^a_\eta\,,  \\ \label{J77}
\ee
\be
k_\nu J^{a}_{\mu\nu}(7,7)=i\frac{g^3_\rho}{g_\omega}
\vep_{\alpha\beta\mu\kappa}k_\kappa q_{1\alpha}
\Delta^\rho_{\beta\eta}(q_1)\Gamma^a_\eta\,,  \\ \label{kJ77}
\ee
\be
q_\mu J^{a}_{\mu\nu}(7,7)=if_\pi M^{a}_\nu(7,5c)\,. \label{qJ77}
\ee
Combining Eq.\,(\ref{qJ75b}) and Eq.\,(\ref{qJ77}) we obtain another PCAC constraint
\be
q_\mu [J^{a}_{\mu\nu}(7,5b)+J^{a}_{\mu\nu}(7,7)]=if_\pi m^2_\pi \Delta^\pi_F(q^2)
M^{a}_\nu(7,5c)\,. \label{qJ75b7}
\ee
From the last vertex $\bar {\cal L}_{7,8}$, one obtains the amplitude
of Fig.\,2i. It reads, together with the divergence, as follows
\be
J^{a}_{\mu\nu}(7,8)=-i \frac{g^3_\rho}{g_\omega}
\vep_{\alpha\beta\mu\nu}q_{1\alpha}
\Delta^\rho_{\beta\eta}(q_1)\Gamma^a_\eta\,,  \\ \label{J78}
\ee
\be
k_\nu J^{a}_{\mu\nu}(7,8)=-i \frac{g^3_\rho}{g_\omega}
\vep_{\alpha\beta\mu\nu}k_\nu q_{1\alpha}
\Delta^\rho_{\beta\eta}(q_1)\Gamma^a_\eta\,,  \\ \label{kJ78}
\ee
\be
q_\mu J^{a}_{\mu\nu}(7,8)
=if_\pi M^{a}_\nu(7,6h) \,.   \label{qJ78}
\ee
It follows from Eq.\,(\ref{kJ77}) and Eq.\,(\ref{kJ78}) that
\be
k_\nu [J^{a}_{\mu\nu}(7,7)+J^{a}_{\mu\nu}(7,8)]=0\,. \label{kJ768}
\ee
From Eq.\,(\ref{qJ76b}) and Eq.\,(\ref{qJ78}) we have the PCAC
constraint
\be
q_\mu [J^{a}_{\mu\nu}(7,6b)+J^{a}_{\mu\nu}(7,8)]=
if_\pi m^2_\pi \Delta^\pi_F(q^2) M^{a}_\nu(7,6h)\,. \label{qJ76b8}
\ee
Let us note that the amplitudes that contain the $\omega$ meson propagator
can be simplified at once, since it holds for the real photon with a good accuracy
\be
\Delta^\omega_{\nu\kappa}(k)\approx \delta_{\nu\kappa}/m^2_\omega\,. \label{OPK20}
\ee

Summing up, the vector--axial-vector (VA) amplitudes are
\begin{center}
$J^{a}_{\mu\nu}(7,1b)$, $J^{a}_{\mu\nu}(7,3)$, $J^{a}_{\mu\nu}(7,5b)$,
$J^{a}_{\mu\nu}(7,6b)$, $J^{a}_{\mu\nu}(7,7)$ and $J^{a}_{\mu\nu}(7,8)$.
\end{center}
Using Eq.\,(\ref{OPK20}), we have from Eq.\,(\ref{M75c}) and Eq.\,(\ref{M76h})
\be
M^{a}_\nu(7,5c)\approx -M^{a}_\nu(7,6h)\,, \label{CM5c6h}
\ee
from which it follows that
\be
J^{a}_{\mu\nu}(7,5b)+J^{a}_{\mu\nu}(7,6b)\approx 0\,.  \label{CJ5c6b}
\ee
It also follows that in this approximation
\be
J^{a}_{\mu\nu}(7,7)+J^{a}_{\mu\nu}(7,8)\approx 0\,.  \label{CJ78}
\ee
Then from the VA amplitudes, only $J^{a}_{\mu\nu}(7,1b)$ and $J^{a}_{\mu\nu}(7,3)$
survive. Since $\Gamma^a_\eta$ depends on the momentum transfer $q_1$,
we can write
\be
J^{a}_{\mu\nu}(7,1b)=-i\frac{g^3_\rho}{g_\omega}
\vep_{\nu\eta\lambda\kappa}k_\kappa q_\lambda q_\mu
\Delta^\pi_F(q^2) \Delta^\rho_F(q^2_1)\Gamma^a_\eta\,,\label{AJ71b}
\ee
\be
J^{a}_{\mu\nu}(7,3)=i\frac{g^3_\rho}{g_\omega}
\vep_{\nu\eta\mu\kappa}k_\kappa\Delta^\rho_F(q^2_1)\Gamma^a_\eta\,.\label{AJ73}
\ee

The vector--vector (VV) amplitudes are
\begin{center}
$J^{a}_{\mu\nu}(7,1a)$, $J^{a}_{\mu\nu}(7,2)$, $J^{a}_{\mu\nu}(7,5a)$
and $J^{a}_{\mu\nu}(7,6a)$.
\end{center}
For the weak momentum transfer $q_\mu$ of interest,
the following approximation is quite acceptable
\be
\Delta^\rho_{\beta\mu}(q)\approx \delta_{\beta\mu}/m^2_\rho\,. \label{RPQ20}
\ee
This approximation simplifies the analysis of the VV amplitudes to a great extent,
since it is clear from Eq.\,(\ref{J71a})  and Eq.\,(\ref{J72}) that
\be
J^{a}_{\mu\nu}(7,1a)+J^{a}_{\mu\nu}(7,2)\approx 0\,, \label{CJ712}
\ee
and from Eq.\,(\ref{J75a}) and Eq.\,(\ref{J76a}) we have
\be
J^{a}_{\mu\nu}(7,5a)+J^{a}_{\mu\nu}(7,6a)\approx 0\,. \label{CJ7a56}
\ee
It means that the VV amplitudes do not contribute to the anomalous
amplitude in the approximation (\ref{RPQ20}).

In the next step, analogous construction of the amplitudes arising from
the Lagrangians $\bar {{\cal L}}_i$, $i=8,\,9,\,10$ is carried out.
It turns out that only the same amplitudes, $J^{a}_{\mu\nu}(7,1b)$ and $J^{a}_{\mu\nu}(7,3)$,
result.
The full anomalous amplitude
can be expressed, using Eq.\,(\ref{Lan}), as
\bea
J^{a,\,an}_{\mu\nu}&=&(\bar{c}_7-\frac{1}{2}\bar{c}_8+\bar{c}_9-\frac{1}{2}\bar{c}_{10})
[J^{a}_{\mu\nu}(7,1b)+J^{a}_{\mu\nu}(7,3)] \nonumber \\
&=&{\tilde c}[J^{a}_{\mu\nu}(7,1b)+J^{a}_{\mu\nu}(7,3)]\,, \label{Jan}
\eea
where
\be
{\tilde c}={\tilde c}_7+{\tilde c}_9=1.85\times 10^{-2}\,. \label{TC79}
\ee
Here we have used Eqs.\,(\ref{bci}) and Eqs.\,(\ref{ctis}).
The amplitude $J^{a,\,an}_{\mu\nu}$ satisfies the continuity
equations
\bea
k_\nu J^{a,\,an}_{\mu\nu}&=&0\,, \\ \label{kJan}
q_\mu J^{a,\,an}_{\mu\nu}&=&if_\pi m^2_\pi \Delta^\pi_F(q^2)
{\tilde c} M^a_\nu(7,1c)\,. \label{qJan}
\eea

We now proceed to calculate the contributions to the form factors
$g_i$ entering the effective Hamiltonian (\ref{H0eff}). For this purpose, we
multiply the amplitudes $J^{a,WZ}_{\mu\nu}$, Eq.\,(\ref{WZC}), and
$J^{a,\,an}_{\mu\nu}$, Eq.\,(\ref{Jan}),
by $\vep_\nu l_\mu$, where the weak lepton current is
$l_\mu=-i{\bar u}_\nu\gamma_\mu(1+\g5)u_\mu$. After the non--relativistic reduction,
we obtain from the Wess--Zumino amplitude $J^{a,WZ}_{\mu\nu}$
\bea
g'_{4,WZ}&=&-\eta\frac{E^2_k}{8\pi^2 f^2_\pi}\frac{\lambda E_k+y E_\nu}
{2m_\mu}\,g^L_P\,,  \\ \label{WZgp4}
g''_{4,WZ}&=&-\eta\frac{E_k E_\nu}{8\pi^2 f^2_\pi}\frac{\lambda E_k+y E_\nu}
{2m_\mu}\,g^L_P\,,  \\ \label{WZgpp4}
g''_{7,WZ}&=&g'_{11,WZ}=\eta\frac{E^2_k}{8\pi^2 f^2_\pi}\frac{\lambda E_\nu}
{2m_\mu}\,g^L_P\,,  \\ \label{WZgpp7}
g''_{10,WZ}&=&g''_{11,WZ}=\eta\frac{E_k}{8\pi^2 f^2_\pi}\frac{\lambda E^2_\nu}
{2m_\mu}\,g^L_P\,.   \label{WZgpp10}
\eea
Here $\eta=m_\mu/2M$, $E_k$($E_\nu$) is the photon (neutrino) energy,
$y={\hat \nu}\cdot{\hat k}$ and the induced pseudoscalar form factor
$g^L_P=2M m_\mu g_A \Delta^\pi_F((q^L)^2)$ depends on the
momentum $q^L=-q_1$. As it has already been noted, this momentum dependence
does not exhibit the enhancement factor for the photon energies at the
high energy tail of the photon spectrum.

Let us calculate next the contributions from the amplitude $J^{a}_{\mu\nu}(7,1b)$.
We obtain
\be
g_{2,1}=-(1+\kappa_V)\eta\frac{g^3_\rho}{g_A g_\omega}\frac{E_k}{2M}
\frac{|\vec k+\vec \nu|^2}{m^2_\rho}\,g^N_P\,,  \label{A1bg2}
\ee
\be
g'_{10,1}=(1+\kappa_V)\eta\frac{g^3_\rho}{g_A g_\omega}(\frac{E_k}{m_\rho})^2
\frac{E_\nu}{2M}\,g^N_P\,, \label{A1bgp10}
\ee
\be
g''_{10,1}=(1+\kappa_V)\eta\frac{g^3_\rho}{g_A g_\omega}(\frac{E_k}{m_\rho})
\frac{E^2_\nu}{2M m_\rho}\,g^N_P\,.  \label{A1bgpp10}
\ee
Here $g^N_P=2M m_\mu g_A \Delta^\pi_F((q^N)^2)$ depends on the momentum
$q^N=\nu-\mu$ and it provides the enhancement factor $\approx\,3$ in the amplitude
at the high energy tail of the photon spectrum. We have also used the approximation
\be
\Delta^\rho_F(q^2_1)\,\approx\,m^{-2}_\rho\,.   \label{Drhoq}
\ee
The last contributions to calculate are from the amplitude $J^{a}_{\mu\nu}(7,3)$
\bea
g_{2,3}&=&(1+\kappa_V)\eta\frac{g^3_\rho}{g_\omega}\frac{E_k(E_k+yE_\nu)}{m^2_\rho}
\,,  \\ \label{A3g2}
g'_{8,3}&=&-g'_{9,3}=g'_{10,3}=-(1+\kappa_V)\eta\frac{g^3_\rho}{g_\omega}
\frac{E_k E_\nu}{m^2_\rho}\,,  \\ \label{A3gp8}
g''_{8,3}&=&-(1+\kappa_V)\eta\frac{g^3_\rho}{g_\omega}
\frac{E^2_k}{m^2_\rho}\,.  \label{A3gpp8}
\eea
These form factors do not contribute to $g_P$. However, the form factor
$g_{2,3}$ is large at the high energy tail of the photon
spectrum. Together with the form factor $g_{2,1}$ of Eq.\,(\ref{A1bg2}),
they are by a factor $\approx\,1/2M$ larger than other form factors $g_{i,1}$
and $g_{j,3}$.
It can be seen that due to the enhancement factor in the form factor $g_{2,1}$,
these two form factors
are  about the same size at the end of the photon spectrum. Indeed, taking
$|g^N_P|\approx 30$, we have $E_k g^N_P/2M\approx 1.5$. Besides,
$|\vec k+\vec \nu|^2\approx E_k(E_k+yE_\nu)\approx E^2_k$.
On the other hand, comparing the form factors $g_{2,1}$ and $g'_{4,WZ}$ we have
at the high energy tail of the photon spectrum
\be
\frac{g_{2,1}}{g'_{4,WZ}}\approx -4\pi^2\frac{1+\kappa_V}{3g_A}
\frac{m_\mu}{M}\frac{g^N_P}{g^L_P}\approx 40\,.  \label{Rg2gp4}
\ee
In deriving this ratio, we used the Kawarabayashi--Suzuki--Fayazuddin--Riazuddin
relation $2 f^2_\pi g^2_\rho=m^2_\rho$. Nevertheless, the
form factors $g_{2,1}$  and $g_{2,3}$ do not enhance sensibly the photon spectrum,
because of the small factor $\tilde c=1.85\times 10^{-2}$, Eq.\,(\ref{TC79}).
The presence of this small factor in the homogenous part of the anomalous Lagrangian
makes its influence approximately equal to the Wess--Zumino term.

The numerical analysis  shows
\cite{TK} that the contribution of the anomalous processes to the singlet and triplet
capture rates for the RMC in hydrogen is $\approx\,0.2$ \%.

Let us briefly discuss the double radiative muon capture (DRMC) in hydrogen
\be
\mu^-\, +\, p\, \longrightarrow \,\nu_\mu\, +\,\gamma\gamma\,+\, n\,.   \label{DRMCp}
\ee
One of the possible processes contributing to this reaction is presented in Fig.\,3.
\\
\begin{center}
  \includegraphics[height=0.25\textheight]{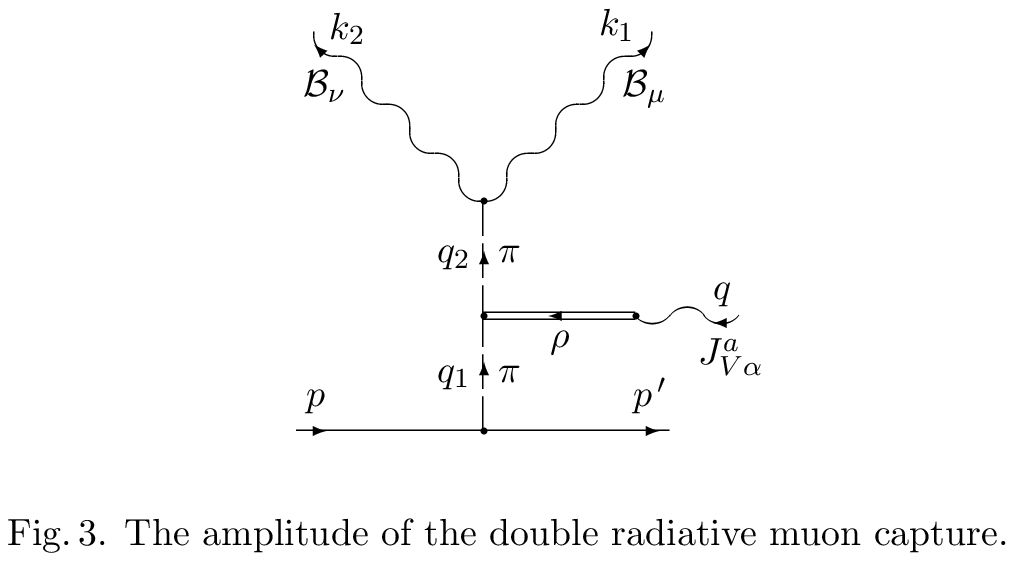}
\end{center}
\vskip 0.35cm

The charged weak vector interaction converts $\pi^+$ into $\pi^0$ that decays into two photons.
This decay is possible due to the chiral anomaly and is described by the anomalous Lagrangian (\ref{Lpi0gg}).
Since in the experiment on the RMC only one photon is detected, the double differential
capture rate for the process (\ref{DRMCp}), integrated over one photon, can contribute
to the measured photon spectrum. Calculations, analogous to those performed above,
provide the following anomalous amplitude for the DRMC process
\be
J^a_{\mu\nu\eta}=\frac{g_{\pi NN}m^2_\rho}{8\pi^2 f_\pi}\vep_{\zeta\nu\alpha\eta}k_{1\zeta}
k_{2\alpha}\Delta^\rho_{\lambda\mu}(q)\Delta^\pi_F(q^2_1)\Delta^\pi_F(q^2_2)(q_{1\lambda}+
q_{2\lambda})\vep^{3ab}\Gamma^b_5\,. \label{JDRMC}
\ee
This amplitude is gauge invariant
\be
k_\nu J^a_{\mu\nu\eta}=k_\eta J^a_{\mu\nu\eta}=0\,.  \label{kkJDRMC}
\ee
In the weak sector it satisfies the following Ward--Takahashi identity
\be
q_\mu J^a_{\mu\nu\eta}=\frac{g_{\pi NN}}{8\pi^2 f_\pi}\vep_{\zeta\nu\alpha\eta}k_{1\zeta}
k_{2\alpha}\left[\Delta^\pi_F(q^2_1)-\Delta^\pi_F(q^2_2)\right]
\vep^{3ab}\Gamma^b_5\,. \label{qJDRMC}
\ee
At the right hand side of Eq.\,(\ref{qJDRMC}), the amplitude of more simple
double radiative process enter. This amplitude is given at Fig.\,1
with the wavy lines corresponding to outgoing photons.

The probability of the DRMC per unit volume is
\be
dw_{fi}=(e^2 G_F \cos\theta_C)^2\frac{1}{2(2\pi)^8}\frac{1}{4}\sum_{s.p.}\,
|M_{fi}|^2\,\delta^4(P_f-P_i)\,
d^3n d^3\nu \frac{d^3k_1}{2E_1}\frac{d^3k_2}{2E_2}\,. \label{dwfi}
\ee
Here
\be
M_{fi}={\tilde l}_\mu(0)\vep^*_\nu(k_1)\vep^*_\eta\,J^a_{\mu\nu\eta}\,. \label{Mfi}
\ee
The form of the DRMC rate resembles that of the single RMC rate
\footnote{For the comparison see, e.g., Eq.\,(4.1) of Ref.\,\cite{TK}.}
\bea  \nonumber
\Lambda_{DRMC}&=&\frac{\alpha^5}{2^8\pi^9}(G_F \cos\theta_C \frac{g_{\pi NN}}{2M f_\pi})^2
m^3 M_n \int \frac{\nu^2_0}{W+E_1(\cos{\theta_1}-1)+E_2(\cos{\theta_2}-1)}
\\ \nonumber
&&\times[\Delta^\pi_F(q^2_1)\Delta^\pi_F(q^2_2)]^2 (k_1\cdot k_2)^2 ({\vec q}_1)^2 F
E_1 d(E_1)E_2  d(E_2) \\ 
&&\times \sin\theta_1 d\theta_1 \sin\theta_2 d\theta_2 d\phi_2
\,.  \label{GDRMC}
\eea
Here
\be
F=[m_\mu-2(E_1+E_2)]^2+4(\vec k_1+\vec k_2)^2-4[m_\mu+2(E_1+E_2)][\hat\nu\cdot(\vec k_1+\vec k_2)]\,,
\label{F}
\ee
\be
\nu_0=\frac{W^2-M^2_n-2W(E_1+E_2)+2E_1 E_2(1-\cos\theta_{12})}{2[W+E_1(\cos\theta_1-1)
+E_2(\cos\theta_2-1)]}\,,  \label{nu0}
\ee
$\alpha$ is the fine structure constant, $W=M_p+m_\mu$, $M_p$ ($M_n$) is
the proton (neutron) mass, $m$ is the
reduced proton-muon mass and $\cos\theta_{12}={{\hat k}_1\cdot{\hat k}_2}$.

In comparison with the single RMC rates \cite{TK}, $\Lambda_{DRMC}$ is suppressed
by the factor $\alpha/2^6\pi^6$. This clearly indicates that $\Lambda_{DRMC}$ is much
smaller. The numerical calculations yield for the triplet capture rate
\be
\Lambda_{DRMC,\,t}\times 10^3\,=\,106.7\times 10^{-13} s^{-1}\,, \label{DMRCt}
\ee
that is damped by the factor $10^{-13}$ in comparison with the
radiative muon capture rate
$\Lambda_t$ for the reaction (\ref{RMCp}) \cite{TK}.

Besides the weak vector amplitude (\ref{JDRMC}), one can also construct
weak axial amplitudes for the DRMC. However, they are expected to provide
an effect of the same order of magnitude.

\section{Conclusions}
\label{CH3}

Starting from the most general anomalous action of the $\pi\rho\omega a_1$
system, we arrive at the 3--point anomalous Lagrangian, that includes
the Wess--Zumino term and four homogenous terms.
Application to radiative muon capture in hydrogen is
considered in order to establish possible contributions
that may enhance the photon spectrum at its high energy tail. It
is important to find and understand the possible sources
that may explain the discrepancy between the experimental
results and the conventional methods to calculate this
spectrum.

The constructed anomalous amplitudes are gauge invariant and in
the weak sector, the vector amplitudes satisfy the CVC hypothesis
and the axial amplitudes satisfy the PCAC constraint.
Using reasonable approximations, only two axial amplitudes
result, besides the Wess--Zumino one. All three terms
yield corrections that
turn out to be of the same order of magnitude for the high photon
energies. The numerical estimates show \cite{TK} that
they cannot provide the necessary enhancement.

We have also made here an estimate of the capture rate for the double
radiative muon capture in hydrogen. This reaction is triggered by the anomalous
decay $\pi^0\rightarrow\gamma\gamma$. The calculations show that
its capture rate turns out to be
strongly suppressed in comparison with the capture rate for the
single radiative muon capture.

Let us note finally that we have recently suggested \cite{TK}
that off--shell effects due to the isobar
intermediate state can, in principle, provide an enhancement
that brings the calculated results in reasonable agreement
with the experiment.

\section*{Acknowledgments}

This work is supported in part by the grant GA \v{C}R 202/03/0210 and by the ASCR project
K1048102. The research of F.~C.~K.~is supported in part by NSERCC.


  \end{document}